\documentclass[letter]{aa}

\usepackage{graphicx}
\usepackage{txfonts}
\usepackage{color}
\usepackage{stfloats}
\usepackage{multirow}
\usepackage{mhchem}
\usepackage{booktabs}
\usepackage{chemformula}
\usepackage{mhchem}
\usepackage[utf8]{inputenc}

\newcommand{\kms}{km\,s$^{-1}$}

\newcommand{\jms}{J. Mol. Spectr.}
\newcommand{\jmst}{J. Mol. Struct.}

\newcommand{\natastro}{Nature Astron.}

\newcommand{\pccp}{PCCP}

\begin{document}

\title{Discovery of interstellar phenalene ($c$-\ch{C13H10}): A new piece for the chemical puzzle of PAHs in space}

\author{
C.~Cabezas\inst{1},
M.~Ag\'undez\inst{1},
C.~P\'erez\inst{2},
D.~Villar-Castro\inst{3},
G.~Molpeceres\inst{1},
D.~P\'erez\inst{3},
A.~L.~Steber\inst{2},
R.~Fuentetaja\inst{1},
B.~Tercero\inst{4,5},
N.~Marcelino\inst{4,5},
A.~Lesarri\inst{2},
P.~de~Vicente\inst{5}, and
J.~Cernicharo\inst{1}
}

\institute{Dept. de Astrof\'{i}sica Molecular, Instituto de F\'{i}sica Fundamental (IFF-CSIC), C/ Serrano 121, 28006 Madrid, Spain\\\email{carlos.cabezas@csic.es, j.cernicharo@csic.es}
\and Departamento de Qu\'imica Física y Qu\'imica Inorg\'anica, Facultad de Ciencias-I.U. CINQUIMA, Universidad de Valladolid, 47011 Valladolid, Spain
\and Centro Singular de Investigaci\'on en Qu\'imica Biol\'oxica e Materiais Moleculares (CiQUS) and Departamento de Qu\'imica Org\'anica, Universidade de Santiago de Compostela, 15782 Santiago de Compostela, Spain
\and Observatorio Astron\'omico Nacional (OAN, IGN), C/ Alfonso XII, 3, 28014, Madrid, Spain
\and Observatorio de Yebes, IGN, Cerro de la Palera s/n, 19141 Yebes, Guadalajara, Spain
}

\date{Received; accepted}

\abstract{We present the discovery of the unsubstituted polycyclic aromatic hydrocarbon (PAH) phenalene (\ch{C13H10}) in TMC-1 as part of the QUIJOTE line survey. In spite of the low dipole moment of this three-ring PAH we have found a total of 267 rotational transitions with quantum numbers $J$ and $K_a$ up to 34 and 14, respectively, corresponding to 100 independent frequencies. The identification of this new PAH from our survey was based on the agreement between the rotational parameters derived from the analysis of the lines and those obtained by quantum chemical calculations. Subsequent chemical synthesis of this PAH and the investigation of its laboratory microwave spectrum unequivocally support our identification. The column density of phenalene in \mbox{TMC-1} is (2.8\,$\pm$\,1.6)$\,\times$\,10$^{13}$ cm$^{-2}$.}


\titlerunning{Discovery of interstellar phenalene}
\authorrunning{Cabezas et al.}

\maketitle

\section{Introduction}

Polycyclic aromatic hydrocarbons (PAHs) are thought to be the most common type of organic molecules in the interstellar medium, based on the strength and ubiquity of the aromatic infrared bands (AIBs), attributed to PAHs forty years ago \citep{Leger1984,Allamandola1985,Tielens2008}. However, no individual PAH has been identified in the interstellar medium (ISM) via the AIBs. The unambiguous proof of the presence of PAHs in the ISM has been provided by radioastronomy through very sensitive line surveys. Two line surveys of the Taurus Molecular Cloud (TMC-1), GOTHAM\footnote{GBT Observations of TMC-1: Hunting Aromatic Molecules} \citep{McGuire2018} and QUIJOTE\footnote{Q-band Ultrasensitive Inspection Journey to the Obscure TMC-1 Environment} \citep{Cernicharo2021a} have unequivocally identify a total of ten particular PAHs; 1- and 2-cyanonaphthalene \citep{McGuire2021}, indene \citep{Cernicharo2021b,Burkhardt2021a}, 2-cyanoindene \citep{Sita2022}, 1- and 5-cyanoacenaphthylene \citep{Cernicharo2024}, 1-, 2- and 4-cyanopyrene \citep{Wenzel2024,Wenzel2025a}, and cyanocoronene \citep{Wenzel2025b}.

With the presence of PAHs confirmed in the ISM, unraveling the formation and subsequent evolution of these interstellar molecules has become a central goal for the PAH community nowadays. The so-called "bottom-up" mechanism has been proposed to explain the formation of PAHs in situ in the cold dense clouds where they are observed \citep{Cernicharo2021c,Byrne2024}. In this scenario, PAHs are built up from small hydrocarbons in the cold and shielded environments of dark clouds. However, the opposite pathway has also been considered, named "top-down" route, where small PAHs result from the fragmentation of very large multi-ringed PAHs inherited from a previous evolutionary stage \citep{Pety2005,Zhen2014,Burkhardt2021b}. Thus, much work remains to be done to fully understand the origin of PAHs in the ISM, starting by obtaining a full inventory of these species and constraining their abundances.

\begin{figure}
\centering
\includegraphics[angle=0,width=0.4\textwidth]{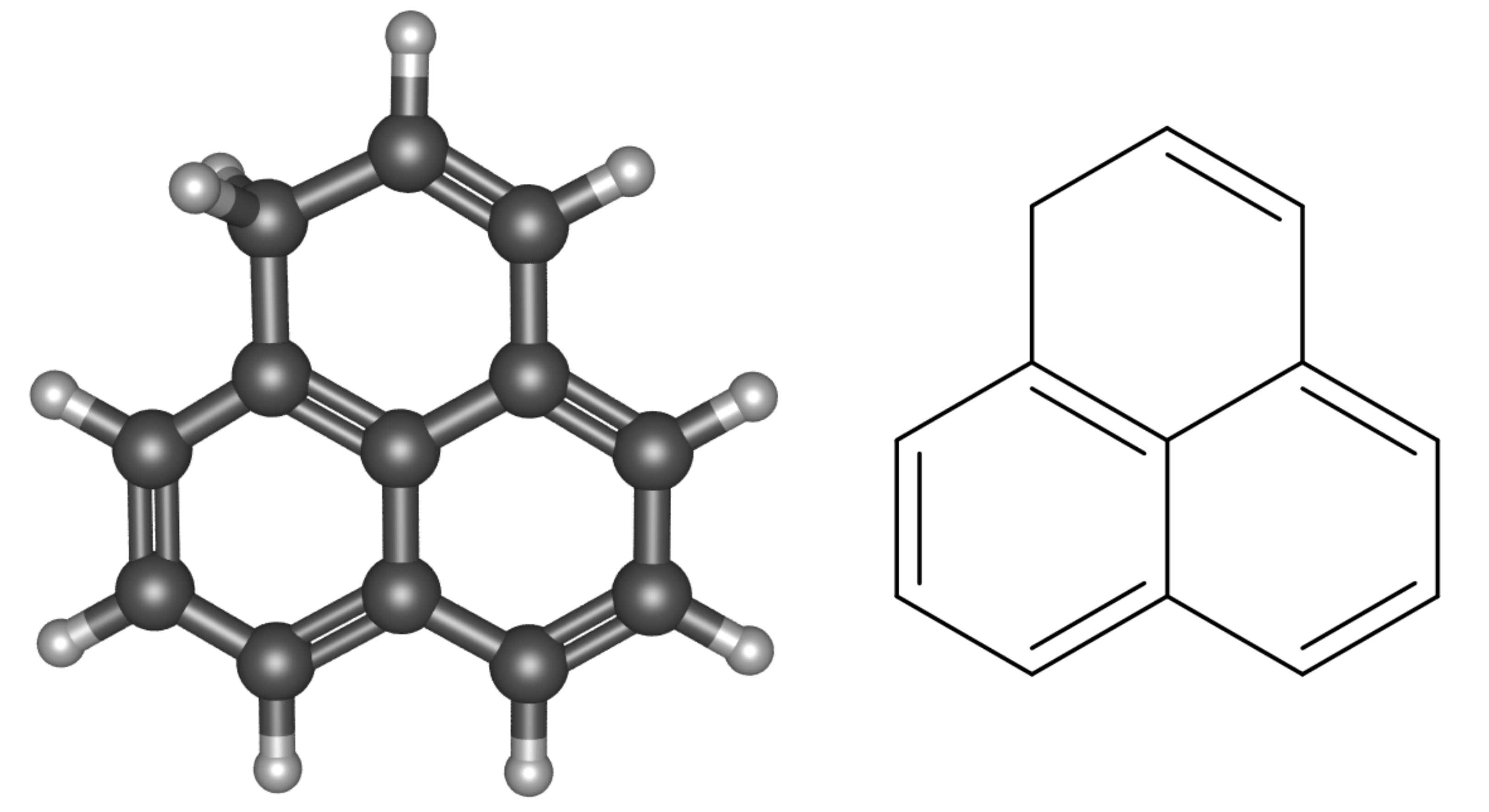}
\caption{3D (left) and 2D (right) molecular structures of phenalene.} \label{molecule}
\end{figure}

From the ten PAHs discovered in TMC-1, only one of them is an unsubstituted PAH, indene, while the other nine species are cyano-functionalized PAHs. This is because PAHs are often highly symmetric or, when they are not symmetric, they are weakly polar at best. This fact makes them notoriously difficult - if not impossible - to detect by their rotational fingerprints. In contrast, replacing a single hydrogen on a pure PAH with a polar functional group, such as the -CN unit, yields a surrogate with a largely increased dipole moment with respect to the non-functionalized PAH counterpart, making it much easier to detect. Nevertheless, one of the most significant results of QUIJOTE is the discovery, through the standard method of line-by-line detection, of low-dipole but very abundant pure hydrocarbons, such as vinyl acetylene (CH$_2$CHCCH; \citealt{Cernicharo2021d}), cyclopentadiene and indene ($c$-C$_5$H$_6$ and $c$-C$_9$H$_8$; \citealt{Cernicharo2021b}), the propargyl radical (H$_2$CCCH; \citealt{Agundez2021,Agundez2022}), and fulvenallene ($c$-C$_5$H$_4$CCH$_2$; \citealt{Cernicharo2022}). In this Letter we report the unequivocal detection of the second unsubstituted PAH found in space, 1$H$-phenalene (phenalene hereafter) with molecular formula \ch{C13H10} (see Fig. \ref{molecule}). This low-dipole moment PAH has been discovered from the QUIJOTE data and its identification has been confirmed by laboratory rotational spectroscopy measurements.

\section{Observations}

The observations of TMC-1 reported here have been acquired with the Yebes 40m telescope as part of the ongoing QUIJOTE project \citep{Cernicharo2021a}. Briefly, QUIJOTE consists of a line survey in the Q band (31.0–50.3 GHz) at the position of the cyanopolyyne peak of TMC-1 ($\alpha_{J2000}=4^{\rm h} 41^{\rm  m} 41.9^{\rm s}$ and $\delta_{J2000}=+25^\circ 41' 27.0''$). This survey was carried out using a receiver built within the Nanocosmos project\footnote{\texttt{https://nanocosmos.iff.csic.es/}} consisting of two cooled high-electron-mobility-transistor (HEMT) amplifiers covering the Q band with horizontal and vertical polarization. Fast Fourier transform spectrometers (FFTSs) with $8\times2.5$ GHz and a spectral resolution of 38.15 kHz provide the whole coverage of the Q-band in both polarizations. Receiver temperatures are 16\,K at 32 GHz and 30\,K at 50 GHz. The experimental setup is described in detail by \citet{Tercero2021}.

All observations are performed in the frequency-switching observing mode with a frequency throw of either 10 or 8 MHz. The total observing time on source for data taken with frequency throws of 10 MHz and 8 MHz is 772.6 and 736.6 hours, respectively. Hence, the total observing time of the QUIJOTE line survey is 1509.2 hours. The sensitivity varies between 0.06 mK at 32 GHz and 0.18 mK at 49.5 GHz. The main beam efficiency can be given across the Q band as $B_{\rm eff}$=0.797 exp[$-$($\nu$(GHz)/71.1)$^2$]. The forward telescope efficiency is 0.97. The telescope beam size at half power intensity is 54.4$''$ at 32.4 GHz and 36.4$''$ at 48.4 GHz. The absolute calibration uncertainty is 10\,$\%$. The data were reduced and processed by using the CLASS package provided within the GILDAS software\footnote{\texttt{http://www.iram.fr/IRAMFR/GILDAS}}. A detailed description of the data-analysis procedure is reported in \citet{Cernicharo2022}.

\section{Results}

\begin{figure*}
\centering
\includegraphics[width=0.80\textwidth]{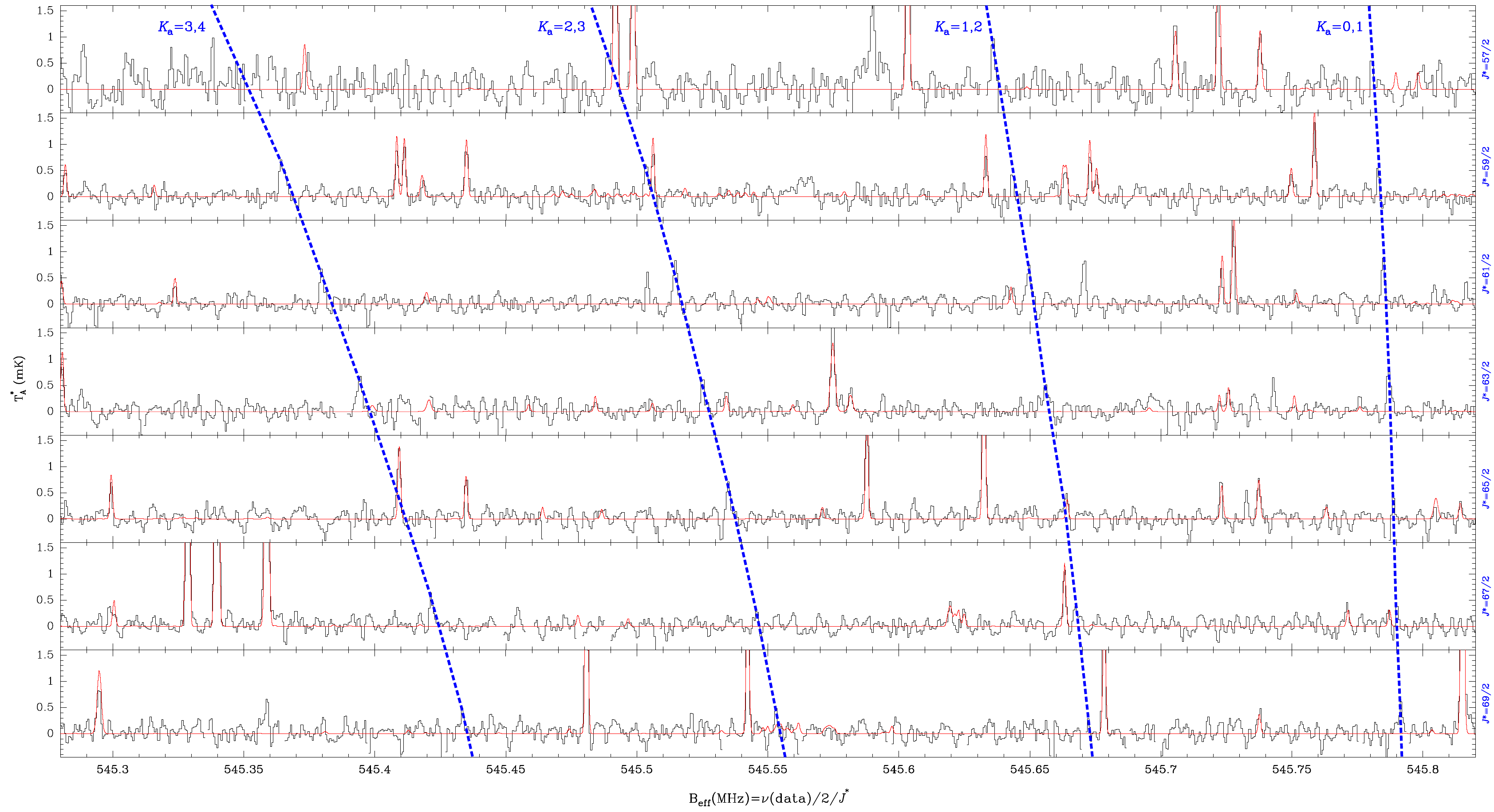}
\caption{Modified Loomis-Wood diagram of some of the observed lines in this work. The abscissa corresponds to the value of the rotational constant, which has been fixed in this plot to values between 545.28 and 545.82 MHz. The ordinate is the antenna temperature ---corrected for atmospheric and telescope losses--- in millikelvin. Each box presents the QUIJOTE data for frequencies 2\,$B_{rot}$\,$J^*$, where $J^*=J_u+1/2$. The red line corresponds to the synthetic spectrum computed for TMC-1, including the rotational transitions of all molecular species detected in TMC-1. This spectrum does not contain the rotational lines discovered in this work. The rotational transitions of phenalene correspond to those connected through dashed blue lines (see text).}
\label{loomis}
\end{figure*}

\begin{table}
\caption{Experimental and theoretical molecular constants of phenalene.} \label{constants}
\centering
\resizebox{9cm}{!}{
\scriptsize
\begin{tabular}{lcccc}
\hline
\hline
Parameter             &         TMC-1        &      Lab          &     TMC-1+Lab   & Theory$^a$   \\
\hline
$A$/MHz               &  1108.1199(12)$^b$ &  1108.12409(26)   & 1108.12404(29)  &  1108.6   \\
$B$/MHz               &   1067.81464(72)   &  1067.81150(22)   & 1067.81175(23)  &  1067.3   \\
$C$/MHz               &    545.85572(15)   &  545.85649(12)    &  545.85613(11)  &   545.6   \\
$\Delta_J$/Hz         &      19.10(63)     &     [19.18]       &     20.21(27)   &   19.18   \\
$\Delta_{JK}$/Hz      &     $-$29.5(13)    &    [$-$29.26]     &  $-$31.30(64)   & $-$29.26  \\
$\Delta_{K}$ /Hz      &      12.20(70)     &     [11.94]       &    13.07(38)    &   11.94   \\
$\delta_J$/Hz         &   [$-$0.237]$^c$   &    [$-$0.237]     &   [$-$0.237]    & $-$0.237  \\
$\delta_K$/Hz         &     [$-$20.56]     &    [$-$20.56]     &   [$-$20.56]    & $-$20.56  \\
$N_t,N_f$$^d$         &       267,71       &      100,55       &      367,126    &    $-$    \\
$\sigma^e$/kHz        &         6.2        &       6.6         &       7.4       &    $-$    \\
$\mu_a$,$\mu_b$/D     &         $-$        &       $-$         &       $-$       &  0.67/0.36  \\
$\kappa$              &        0.86        &       0.86        &       0.86      &   0.85    \\
$\Delta_c$$^f$/amu$Å^2$ &    $-$3.5051(9)    &   $-$3.5060(4)    &   $-$3.5054(4)  &  $-$3.1   \\
\hline
\hline
\end{tabular}
}
\tablefoot{
\tablefoottext{a}{The values for the rotational constants $A$, $B$, and $C$ calculated theoretically correspond to the equilibrium rotational constants $A_e$, $B_e$, and $C_e$.}
\tablefoottext{b}{The uncertainties (in parentheses) are in units of the last significant digits.}
\tablefoottext{c}{Values in brackets were not determinable and were thus fixed to the theoretically calculated ones.}
\tablefoottext{d}{$N_t$ and $N_f$ refer to the number of distinct rotational transitions and unique transition frequencies, respectively, in the fit.}
\tablefoottext{e}{Standard root mean square deviation of the fit.}
\tablefoottext{f}{Inertial defect, $\Delta_c$,=$I_c$-$I_b$-$I_a$. The conversion factor is 505379.1 MHz amu $Å^2$.}\\
}
\end{table}

Line identification in QUIJOTE is performed using the MADEX code \citep{Cernicharo2012}, in conjunction with the CDMS and JPL catalogues \citep{Muller2005,Pickett1998}. The intensity scale used in this study is the antenna temperature ($T_A^*$). Consequently, the telescope parameters and source properties were used when modelling the emission of the different species in order to produce synthetic spectra on this temperature scale. In this work, we assumed a velocity for the source relative to the local standard at rest of 5.83 \kms\, \citep{Cernicharo2020}. The source was assumed to be circular with a uniform brightness temperature and a radius of 40$''$ \citep{Fosse2001,Cernicharo2023}.

Using the same methodology employed to identify the rotational transitions of the cyano derivatives of acenaphthylene \citep{Cernicharo2024}, we found several series of lines among the unknown features in the QUIJOTE line survey (see Fig. \ref{loomis}). We first assigned the lines of the series at higher frequency (right side) to the $a$-type rotational transitions (with $K_a$\,=\,0 and 1) $J$+1$_{0,J+1}$$\leftarrow$$J_{0,J}$ and $J$+1$_{1,J+1}$$\leftarrow$$J_{1,J}$ of a new molecular species. Then, in an iterative fitting procedure, we assigned the second series (from right to left) to the transitions with $K_a$\,=\,1 and 2 $J$+1$_{1,J}$$\leftarrow$$J_{1,J-1}$ and $J$+1$_{2,J}$$\leftarrow$$J_{2,J-1}$. An initial fit was performed using more lines of our survey, with $J$ and $K_a$ values in the ranges 14-33 and 0-14, respectively. The lines were fitted using the SPFIT program \citep{Pickett1991} with the $A$-reduction of the Watson's Hamiltonian and $III^l$ representation \citep{Watson1977}. From this fit we obtained a preliminary set of molecular constants ($A$=1108.11850, $B$=1067.81704 and $C$=545.855894 MHz).

From this set of rotational constants we can infer some information about the nature of the carrier of the observed lines. First, the Ray's parameter, $\kappa$, defined as (2$B$-$A$-$C$)/($A$-$C$), is a measuring of the degree of asymmetry of the molecule. For our carrier, the $\kappa$ value is 0.86, which means that the molecule is a fairly oblate asymmetric rotor. Secondly, the inertial defect $\Delta_c$, defined as $I_c$-$I_b$-$I_a$ ($I_x$ is the inertial moment in the $x$ molecular axes), is $-$3.5051 amu $Å^2$. This value is typical for molecules with a planar heavy atom skeleton and two methylenic hydrogen atoms as the sole out-of-plane mass contributors \citep{Gordy1984}. And finally, the values of the rotational constants point to a molecule with a molecular size a bit larger than that of the PAHs acenaphthene (\ch{C12H10}) and acenaphthylene (\ch{C12H8}) \citep{Thorwirth2007}. In principle, we can discard a cyano derivative of a PAH since the presence of a cyano moiety in a PAH induces a prolate asymmetry in the molecular framework. Examples of some cyano-PAHs with negative values of the $\kappa$ parameter are 2-cyanonaphthalene \citep{McNaughton2018}, 2-cyanopyrene \citep{Wenzel2024} and 1-cyanoacenaphthylene \citep{Cernicharo2024}, with $\kappa$\,=\,$-$0.90, $-$0.81 and $-$0.46, respectively. Based on all these points, the first candidate to be considered as carrier of our lines is a PAH with one more carbon atom than acenaphthene and acenaphthylene. The first possibility we considered was a PAH with a molecular formula \ch{C13H10}, which corresponds to the two isomers fluorene and phenalene, also known as $9H$-fluorene and 1$H$-phenalene. Fluorene can be discarded because its rotational spectrum is already known \citep{Thorwirth2007}. In contrast its isomer phenalene (see structure in Fig. \ref{molecule}), has never been observed in the laboratory. It is a promising candidate since its molecular structure looks to some extent oblate and with an out-of-plane methylene \ce{-CH2} moiety.

To get some insights about the rotational constants of phenalene, we carried out quantum chemical calculations to optimize the molecular structure at the B3LYP/6-311++G(d,p) level of theory \citep{Becke1993,Frisch1984} using the \textsc{Gaussian16} package \citep{Frisch2016}. From these calculations we obtained the values of the rotational constants, centrifugal distortion constants and electric dipole moment components shown in Table \ref{constants}. They agree very well with those derived from the TMC-1 data and point out that the carrier of the lines observed in TMC-1 is the PAH phenalene. The discrepancies between the experimental and calculated values of the rotational constants are smaller than 0.05\%. The calculated electric dipole moment components indicate that the $b$-type transitions also contribute to the observed transitions. In the initial fit we did not consider the $b$-type transitions, but taking into account the predicted electric dipole moment components, we made a new fit of the TMC-1 data including these transitions. In this manner, the rotational spectrum of phenalene in the Q band is such that up to four transitions, two $a$- and two $b$-type with the appropriate $K_a$ and $K_a$+1 values (see Fig.\ref{tmc}), collapse to the same frequency. The final dataset of phenalene in TMC-1 includes 71 frequencies and 267 transitions. A list of the assigned transitions with the line parameters is given in Table\,\ref{table:line_parameters}. The final fit for the lines observed in TMC-1 provides the set of molecular constants, rotational and centrifugal, shown in the left column of Table \ref{constants}. The agreement between the derived constants and those obtained from quantum chemical calculations indicate that phenalene is the carrier of the series of lines observed in TMC-1.

In order to fully confirm the astronomical discovery of phenalene in TMC-1, we carried out laboratory experiments to observe its microwave spectrum. The commercial availability of phenalene is limited, and we thus synthesized it in the laboratory. We prepared phenalene in three steps from 3-(naphthalen-1-yl)propanoic acid (see Fig. \ref{schemes}), through a minor adaptation of previously published procedures \citep{Zhao2020,Turco2023}, see details in Appendix \ref{synthesis}, and measured its rotational spectrum from 2-11 GHz using a broadband chirped-pulse Fourier-transform microwave (CPFTMW) spectrometer, see details in Appendix \ref{ftmw}. In the laboratory spectrum of phenalene we measured 55 rotational frequencies that correspond to a total of 100 rotational transitions, including $a$- and $b$-type ones. All the observed frequencies, which are available on Zenodo repository (see Sect.\,\ref{data}), were fitted using the SPFIT program \citep{Pickett1991} with the $A$-reduction of the Watson's Hamiltonian and $III^l$ representation \citep{Watson1977}. We only fit the values of the rotational constants, while the centrifugal distortion constants were kept fixed to the theoretical values since not enough accurate values could be derived with the available data. The values obtained from the analysis are shown in Table \ref{constants}. A combined fit of the laboratory and TMC-1 data for phenalene provides the recommended molecular parameters, which are given in Table \ref{constants}. As it can be seen, there is no doubt that phenalene is the carrier of the lines observed in TMC-1. Hence, we firmly conclude that we have detected in TMC-1 the second unsubstituted PAH after indene, detected in 2021 \citep{Cernicharo2021d}.

\section{Discussion}\label{Discussion}

In order to estimate the abundance of phenalene in TMC-1, we constructed a rotation diagram, assuming that it is distributed as a circle with a radius of 40$''$ as benzonitrile \citep{Cernicharo2023}. We obtained a rotational temperature of 7.9\,$\pm$\,1.2 K, which is not far from the gas kinetic temperature of 9 K in TMC-1 \citep{Agundez2023}, and a column density of (2.8\,$\pm$\,1.6)$\,\times$\,10$^{13}$ cm$^{-2}$. It should be noted that these values have been estimated using the theoretical dipole moment components from Table \ref{constants}. The calculated line profiles are compared with the observed ones in Fig. \ref{tmc}. The column density of phenalene is somewhat larger than that of the other pure PAH detected in TMC-1, indene, for which \cite{Cernicharo2021b} derived a column density of 1.6$\,\times$\,10$^{13}$ cm$^{-2}$. It is also interesting to note that the column density of these two pure PAHs is within the range of column densities inferred for the PAHs naphthalene, acenaphthylene, pyrene, and coronene, (1-10)\,$\times$\,10$^{13}$ cm$^{-2}$ \citep{McGuire2021,Cernicharo2024,Wenzel2024,Wenzel2025a,Wenzel2025b}. We, however, note that the latter range has a significant uncertainty due to the conversion factor used to scale the abundance of the cyano derivative to that of the corresponding unsubstituted PAHs, where factors in the range 5-30 are typically estimated \citep{Cernicharo2022,Wenzel2025b,Steber2025}.

It is puzzling how phenalene and the other PAHs detected in \mbox{TMC-1} can be formed with such large abundances, in the range 10$^{-9}$-10$^{-8}$ relative to H$_2$. If PAHs are formed through a bottom-up mechanism, we can think of possible chemical reactions able to synthesize phenalene from smaller cycles. Neutral-neutral reactions between smaller PAHs and reactive radicals could in principle do the job. If exothermic and barrierless, these reactions commonly occur with H atom elimination. There are several potentially interesting reactions, such as the reactions between naphthalene and propargyl radical (\ch{C10H8} + \ch{C3H3}), indene (\ch{C9H8}) and the \ch{C4H3} radical, and acenaphthylene (\ch{C12H8}) and the \ce{CH3} radical, all of which involve reactants which are known or expected to be abundant in \mbox{TMC-1}. However, these reactions are calculated to have barriers in the channel yielding phenalene, as we show using our own quantum chemical calculations (Appendix \ref{sec:qm}). Another bottom-up route could be provided by ion-neutral reactions able to form protonated phenalene (\ch{C13H11+}), the dissociative recombination of which with electrons could yield phenalene. An investigation of the dissociation channels of \ch{C13H11+} using automatic reaction discovery methods \citep{maeda_afir_2013,maeda_toward_2023} shows as a promising pathway the reaction of acenaphthylene \ce{C12H8} with the methyl cation \ce{CH3+}, as an exothermic and barrierless reaction (Appendix \ref{sec:qm}), following the emission of a photon, i.e, \ce{C12H8 + CH3+ -> C13H11+ + h$\nu$} and a subsequent dissociative recombination, as mentioned above to form phenalene. This highlights the importance of reconsidering ion-neutral reactions, including radiative associations in the formation of PAHs in the ISM.

\section{Conclusions}

In the present work, we report the detection of the second unsubstituted PAH, phenalene, in TMC-1. The discovery of this PAH is based on the observation of 71 lines in our line survey QUIJOTE and supported by laboratory rotational spectroscopy experiments. The derived column density of phenalene is similar to those found for other PAHs in the same source. The chemistry of PAHs in the ISM is still not understood, specially for medium-sized PAHs such as phenalene. Our preliminary results for this molecule suggest that ion-molecule chemistry might have a greater role than expected.

\section{Data availability}
\label{data}

Data underlying this article are available on Zenodo repository. The data comprise the line parameters toward TMC-1 and the laboratory measured frequencies for phenalene.

\begin{acknowledgements}

This work was based on observations carried out with the Yebes 40m telescope (projects 19A003, 20A014, 20D023, 21A011, 21D005, and 23A024). The 40m radiotelescope at Yebes Observatory is operated by the IGN, Ministerio de Transportes y Movilidad Sostenible. We acknowledge funding support from Spanish Ministerio de Ciencia e Innovación through grants PID2019-106110GB-I00, PID2019-106235GB-I00, PID2021-125015NB-I00, PID2022-137980NB-I00, PID2022-139933NB-I00, and PID2023-147545NB-I00. C.P. thanks the ERC for the CoG HydroChiral (Grant Agreement No 101124939). C.P., A.L.S. and A.L. also thank funding from Junta de Castilla y León, Grant INFRARED IR3032-UVA13. D.V. and D.P. thank Xunta de Galicia and ERDF for grant ED431G 2023/03. G.M. and A.L.S. acknowledges the support of the grants RYC2022-035442-I and RYC2022-037922-I, respectively, funded by MICIU/AEI/10.13039/501100011033 and ESF+. G.M. also received support from project 20245AT016 (Proyectos Intramurales CSIC).

\end{acknowledgements}

\begin{appendix}

\section{Line parameters of phenalene}\label{line_parameters}

\begin{figure*}
\centering
\includegraphics[angle=0,width=0.9\textwidth]{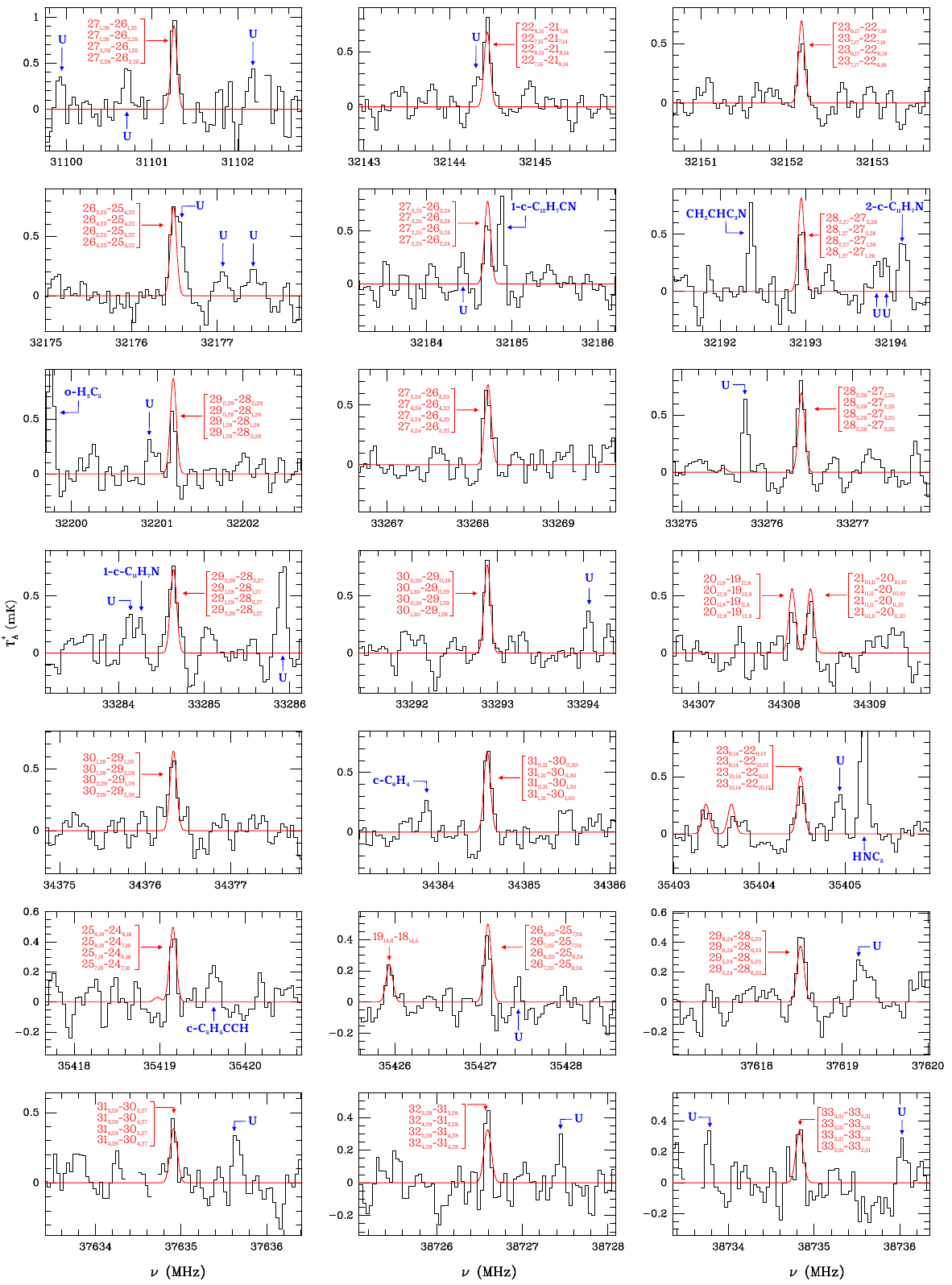}
\caption{A selection of observed lines of phenalene in TMC-1 from the QUIJOTE line survey (black histogram) and synthetic spectra (red curve) calculated adopting a column density of 2.8$\times$10$^{13}$ cm$^{-2}$. Blanked channels correspond to negative features produced in the folding of the frequency switching data. The lines are indicated by arrows and their quantum numbers are depicted for each feature. Other lines from molecules previously studied or unknown features are indicated in blue. The abscissa corresponds to the rest frequency assuming a local standard of rest velocity of 5.83\,\kms. The ordinate is the antenna temperature in millikelvin.} \label{tmc}
\end{figure*}

Line parameters for all observed transitions with the Yebes 40m radio telescope were derived by fitting a Gaussian line profile to them using the GILDAS package. A velocity range of $\pm$50\,\kms\, around each feature was considered for the fit after a polynomial baseline was removed. Negative features in the folding of the frequency switching data were blanked
before baseline removal. A view of some of the transitions of phenalene is given in Fig. \ref{tmc}.

\begin{table}
\caption{Observed line parameters of phenalene in TMC-1.$^a$}\label{table:line_parameters}
\centering
\scriptsize
\begin{tabular}{lcccc}
\hline
\hline
Transition$^b$ &$\nu_{obs}$$^c$  & $\int$ $T_A^*$ dv $^d$ & $\Delta$v$^e$   &   $T_A^*$$^f$\\
           &      (MHz)      &  (mK\,km\, s$^{-1}$)    &  (km s$^{-1}$)  &   (mK)       \\
\hline
$27_{\beta,26}$-$26_{\beta,25}$          &  31101.252$\pm$0.010 &  0.90$\pm$0.32 &  0.81$\pm$0.35 & 1.03$\pm$0.21\\
$28_{\alpha,28}$-$27_{\alpha,27}$        &  31109.488$\pm$0.010 &  0.39$\pm$0.19 &  0.51$\pm$0.24 & 0.70$\pm$0.19\\
$19_{\lambda,9}$-$18_{\lambda,8}$        &  32129.188$\pm$0.010 &  0.33$\pm$0.10 &  0.62$\pm$0.30 & 0.50$\pm$0.11\\
$21_{\iota,13}$-$20_{\iota,12}$          &  32137.269$\pm$0.010 &  0.36$\pm$0.12 &  0.72$\pm$0.24 & 0.47$\pm$0.09\\
$22_{\theta,15}$-$21_{\theta,14}$        &  32144.436$\pm$0.010 &  0.91$\pm$0.11 &  1.20$\pm$0.21 & 0.71$\pm$0.09\\
$23_{\eta,17}$-$22_{\eta,16}$            &  32152.173$\pm$0.010 &  0.51$\pm$0.11 &  0.90$\pm$0.21 & 0.53$\pm$0.08\\
$24_{\zeta,19}$-$23_{\zeta,18}$          &  32160.171$\pm$0.010 &  0.41$\pm$0.12 &  0.93$\pm$0.37 & 0.41$\pm$0.07\\
$25_{\epsilon,21}$-$24_{\epsilon,20}$    &  32168.288$\pm$0.010 &  0.35$\pm$0.10 &  0.71$\pm$0.22 & 0.46$\pm$0.08\\
$26_{\delta,23}$-$25_{\delta,22}$        &  32176.490$\pm$0.010 &  0.44$\pm$0.07 &  0.76$\pm$0.15 & 0.53$\pm$0.11\\
$27_{\gamma,25}$-$26_{\gamma,24}$        &  32184.712$\pm$0.010 &  0.49$\pm$0.09 &  0.74$\pm$0.15 & 0.62$\pm$0.12\\
$28_{\beta,27}$-$27_{\beta,26}$          &  32192.950$\pm$0.010 &  0.45$\pm$0.12 &  0.69$\pm$0.21 & 0.60$\pm$0.10\\
$29_{\alpha,29}$-$28_{\alpha,28}$        &  32201.186$\pm$0.010 &  0.36$\pm$0.12 &  0.51$\pm$0.28 & 0.67$\pm$0.09\\
$21_{\kappa,12}$-$20_{\kappa,11}$        &  33222.247$\pm$0.010 &  0.26$\pm$0.07 &  0.58$\pm$0.18 & 0.42$\pm$0.10\\
$22_{\iota,14}$-$21_{\iota,13}$          &  33228.623$\pm$0.010 &  0.56$\pm$0.11 &  0.95$\pm$0.21 & 0.56$\pm$0.08\\
$27_{\delta,24}$-$26_{\delta,23}$        &  33268.165$\pm$0.010 &  0.55$\pm$0.09 &  0.79$\pm$0.15 & 0.66$\pm$0.09\\
$28_{\gamma,26}$-$27_{\gamma,25}$        &  33276.398$\pm$0.010 &  0.79$\pm$0.09 &  0.93$\pm$0.12 & 0.80$\pm$0.10\\
\hline
\hline
\end{tabular}
\tablefoot{
\tablefoottext{a}{The full content of this table can be found in electronic form at Zenodo (see Section \ref{data}). All uncertainties correspond to 1$\sigma$.}
\tablefoottext{b}{Each frequency of second column is assigned to four rotational transitions, two $a$- and two $b$-type ones. Labels $\alpha, \beta, \gamma, \delta, \epsilon, \zeta, \eta, \theta, \iota, \kappa, \lambda$, correspond to $K_a$=0,1, $K_a$=1,2, $K_a$=2,3, $K_a$=3,4, $K_a$=4,5, $K_a$=5,6, $K_a$=6,7, $K_a$=7,8, $K_a$=8,9, $K_a$=9,10, and $K_a$=10,11, respectively, for each quartet of lines. As an example, the line at 31101.252 MHz is assigned to the quartet of lines $27_{\beta,26}$-$26_{\beta,25}$, which means to $27_{0,26}$-$26_{0,25}$, $27_{0,26}$-$26_{1,25}$, $27_{1,26}$-$26_{0,25}$ and $27_{1,26}$-$26_{1,25}$ transitions. }
\tablefoottext{c}{Measured frequency assuming a v$_{LSR}$ of 5.83 km\,s$^{-1}$ for TMC-1 \citep{Cernicharo2020}.}\\
\tablefoottext{d}{Integrated line intensity (in mK\,km\,s$^{-1}$).}
\tablefoottext{e}{Linewidth at half intensity derived by fitting a Gaussian function to the observed line profile (in km\,s$^{-1}$).}
\tablefoottext{f}{Antenna temperature (in milli Kelvin).}
}
\end{table}

\section{Synthesis of phenalene}\label{synthesis}

\subsection{General synthetic methods}

\begin{figure}[]
\centering
\includegraphics[width=0.40\textwidth]{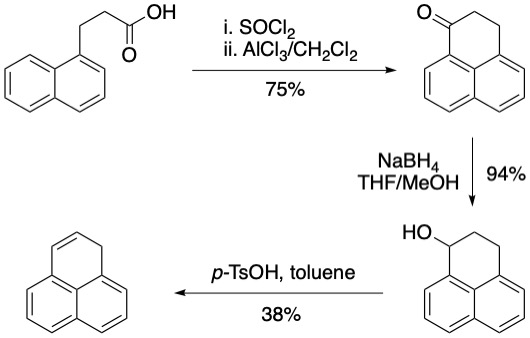}
\caption{Synthesis 1$H$-phenalene.}
\label{schemes}
\end{figure}

Reactions were carried out under argon using oven-dried glassware. THF and \ch{CH2Cl2} were obtained from a MSBraun SPS-800  Solvent Purification System. Other commercial reagents were purchased from Sigma-Aldrich or BLD-Pharm and used without further purification. 1$H$-phenalene was synthetized (Fig. \ref{schemes}) following reported procedures with minor modifications \citep{Zhao2020,Turco2023}. Thin-Layer Comatography (TLC) was performed on Merck silica gel 60 F254 and chromatograms were visualized with UV Light (254 nm and 360 nm). Column cromatography was performed on Merck silica gel 60 (ASTM 40-60 $\mu$m). NMR spectra were recorded at 500 MHz and 125 MHz for $^1$H and $^{13}$C respectively (Bruker DRX-500). Gas Cromatography/Mass Spectrometry (GC/MS) analysis were conducted on a HP 5973 INERT coupled to Agilent HP-5MS. Atmospheric pressure chemical ionization (APCI) High Resolution Mass Spectrometry spectra were obtained on a Bruker Microtof using Direct Inlet Probe (DIP) for sample introduction.

\subsection{Experimental procedures and characterization data}

\subsubsection{Synthesis of 2,3-dihydro-1$H$-phenalen-1-one}

A mixture of 3-(naphthalen-1-yl)propanoic acid (506 mg, 2.42 mmol) and \ch{SOCl2} (4 mL) was heated to 80$^{\circ}$C for 1 h. The volatiles were removed under vacuum and the resulting oily crude was dissolved in dry \ch{CH2Cl2} (2 mL) and it was added dropwise to a solution of \ch{AlCl3} (759 mg, 5.70 mmol) in dry \ch{CH2Cl2} (9 mL) at -20$^{\circ}$C. The mixture was stirred for 30 min and it was poured into ice. Layers were separated and the aqueous phase was extracted with \ch{CH2Cl2}. The organics were dried over anhydrous \ch{Na2SO4}, the solvent was removed by rotary evaporation and the crude was purified by column cromatography (\ch{SiO2}, hexane/AcOEt 4:1) to obtain 2,3-dihydro-1$H$-phenalen-1-one as a colorless solid (332 mg, 1.82 mmol, 75\%). Spectroscopic data were identical to those reported in the literature \citep{Turco2023}.

\begin{figure}[]
\centering
\includegraphics[width=0.45\textwidth]{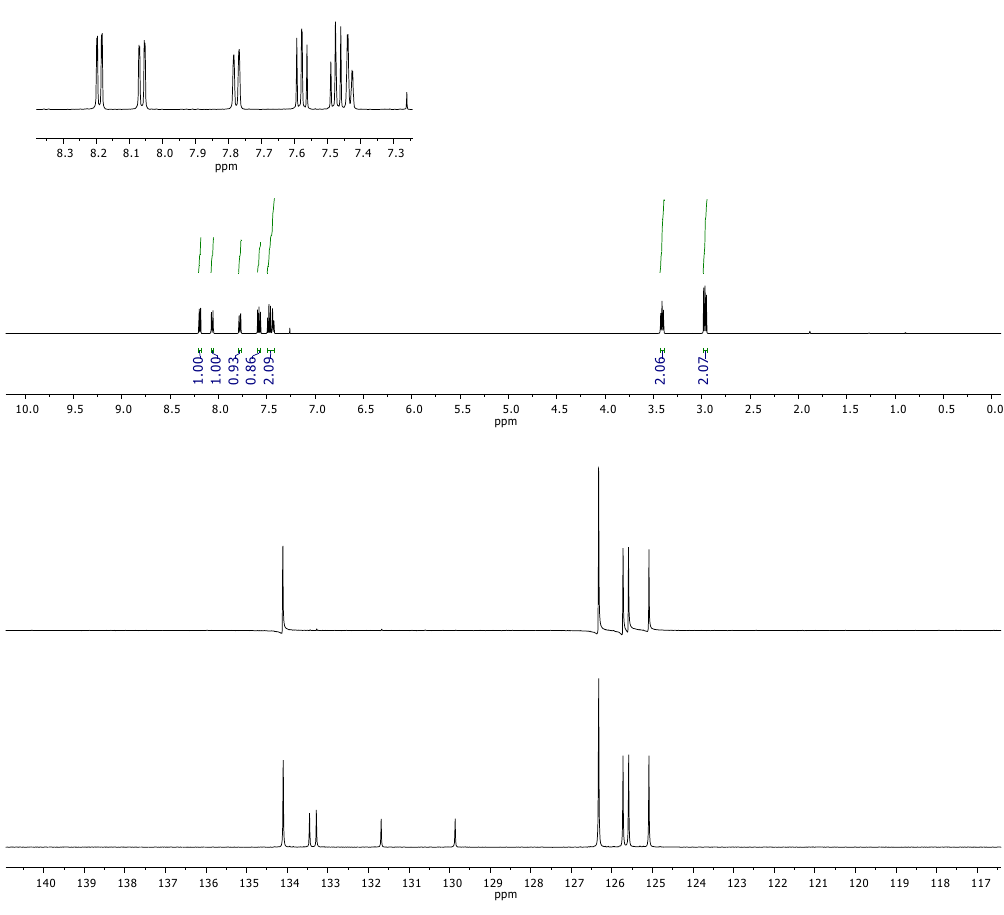}
\caption{$^1$H and $^{13}$C-NMR spectra of 2,3-dihydro-1$H$-phenalen-1-one.}
\label{spect_1}
\end{figure}

\textbf{$^1$H-NMR} (500 MHz, Chloroform-$d$) $\delta$: 8.19 (dd, $J$ = 7.2, 1.3 Hz, 1H), 8.06 (dd, $J$ = 8.1, 1.3 Hz, 1H), 7.78 (dd, $J$ = 8.3, 1.2 Hz, 1H), 7.58 (dd, $J$ = 8.2, 7.2 Hz, 1H), 7.51 – 7.40 (m, 2H), 3.41 (m, 2H), 2.99-2.94 (m, 2H). \textbf{$^{13}$C-NMR-DEPT} (126 MHz, Chloroform-$d$), $\delta$: 198.54 (C), 134.10 (CH), 133.45 (C), 133.28 (C), 131.68 (C), 129.86 (C), 126.32 (2xCH), 125.72 (CH), 125.59 (CH), 125.09 (CH), 38.56 (CH$_2$), 28.60 (CH$_2$) ppm. Spectra shown in Fig. \ref{spect_1}.

\subsubsection{Synthesis of 2,3-dihydro-1$H$-phenalen-1-ol}

2,3-dihydro-1$H$-phenalen-1-one (311 mg, 1.82 mmol) was dissolved in a mixture 5:3 of dry MeOH/THF (8 mL) and \ch{NaBH4} (137 mg, 3.64 mmol) was added at 0$^{\circ}$C in one portion. The mixture was stirred for 5 min and then it was allowed to warm up to room temperature and stirred for 30 min. Water and AcOEt were added, layers were separated and the aqueous phase was extracted with AcOEt. The organics were dried over anhydrous \ch{Na2SO4} and solvents were removed by rotary evaporation to afford pure 2,3-dihydro-1$H$-phenalen-1-ol as a colorless solid (314 mg, 1.70 mmol, 94\%). Spectroscopic data were identical to those reported in the literature \citep{Turco2023}.

\begin{figure}[]
\centering
\includegraphics[width=0.450\textwidth]{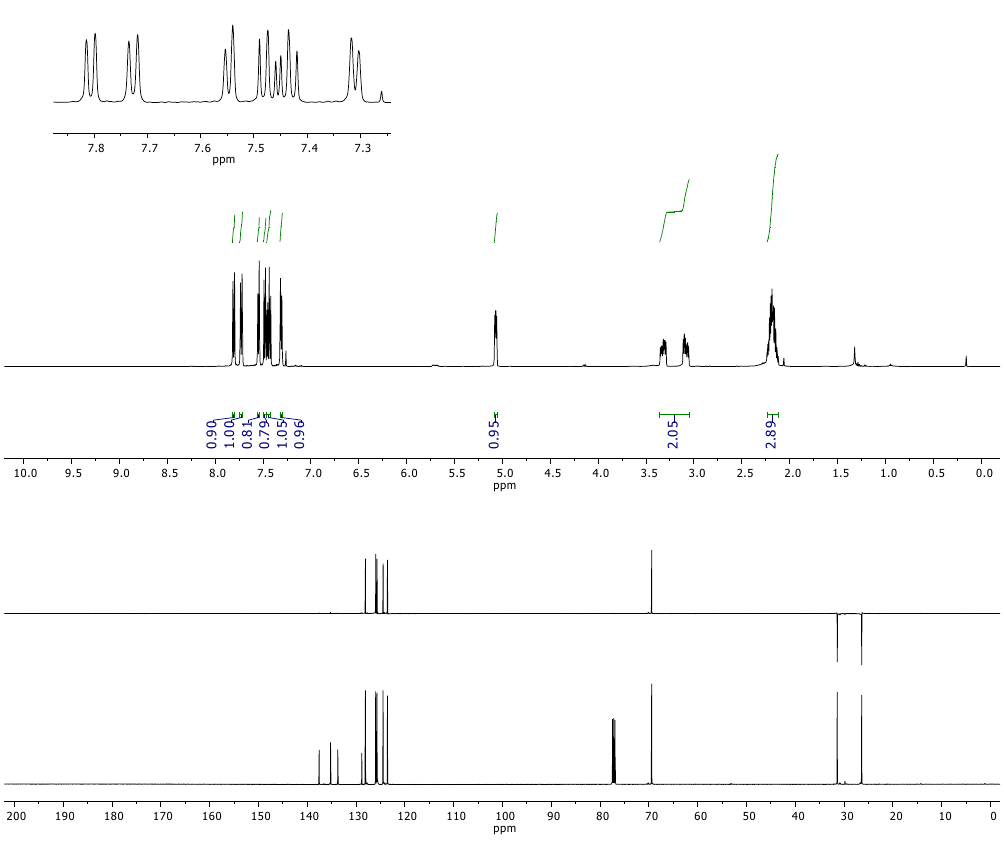}
\caption{$^1$H and $^{13}$C-NMR spectra of 2,3-dihydro-1$H$-phenalen-1-ol.}
\label{spect_2}
\end{figure}

\textbf{$^1$H-NMR} (500 MHz, Chloroform-$d$) $\delta$: 7.81 (d, $J$ = 8.2 Hz, 1H) 7.73 (d, $J$ = 8.2 Hz, 1H), 7.55 (d, $J$ = 7.0 Hz, 1H), 7.48 (d, $J$ = 7.6 Hz, 1H), 7.46-7.42 (m, 1H), 7.31 (d, $J$ = 7.0 Hz, 1H), 5.07 (dd, $J$ = 6.8, 3.8 Hz, 1H), 3.40-3.04 (m, 2H), 2.18 (m, 2H).\textbf{$^{13}$C-NMR-DEPT} (126 MHz, Chloroform-$d$), $\delta$: 137.53 (C), 135.19 (C), 133.71 (C), 128.79 (C), 128.05 (CH), 125.96 (CH), 125.73 (CH), 125.66 (CH), 124.43 (CH), 123.54 (CH), 69.40 (CH), 31.39 (CH$_2$), 26.37 (CH$_2$) ppm. Spectra shown in Fig. \ref{spect_2}.

\subsubsection{Synthesis of 1$H$-phenalene}

A catalytic amount of p-TsOH was added to a solution of 2,3-dihydro-1$H$-phenalen-1-ol (314 mg, 1.70 mmol) in toluene (10 mL) with 4 Å sieves and the mixture was heated to 110$^{\circ}$C for 30 min. Then it was cooled to room temperature and the solvent was removed. The crude was purified by flash column chromatography (SiO$_2$, hexane) to obtain 1$H$-phenalene as a colorless solid (108 mg, 0.65 mmol, 38\%). Spectroscopic data were identical to those reported in the literature \citep{Zhao2020,Turco2023}.

\begin{figure}[]
\centering
\includegraphics[width=0.450\textwidth]{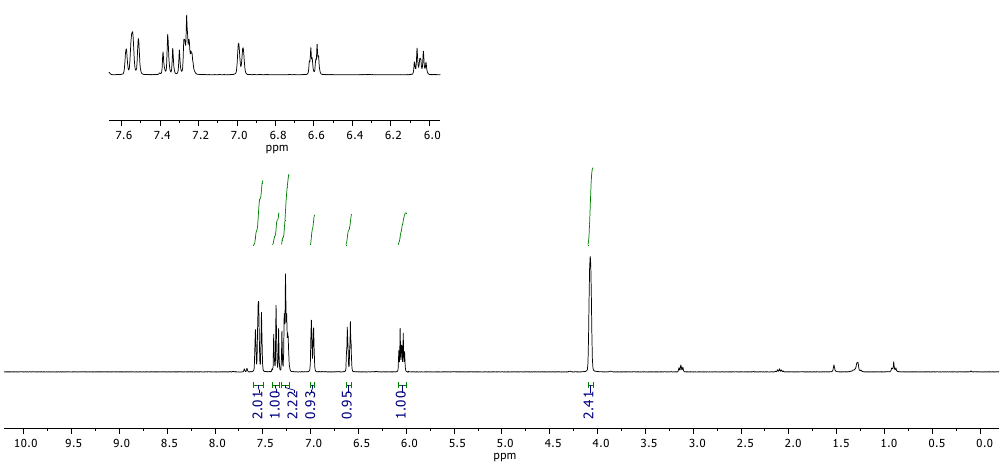}
\caption{$^1$H spectrum of 1$H$-phenalene.}
\label{spect_3}
\end{figure}

\textbf{$^1$H-NMR} (300 MHz, Chloroform-$d$) $\delta$: 7.59 – 7.50 (m, 2H), 7.36 (t, $J$ = 7.6 Hz, 1H), 7.31 – 7.22 (m, 2H), 6.98 (d, $J$ = 6.9 Hz, 1H), 6.60 (dt, $J$ = 9.9, 2.3 Hz, 1H), 6.05 (dt, $J$ = 9.9, 4.1 Hz, 1H), 4.08 (bs, 2H). Mp 80-81$^{\circ}$C (Lit. 83-84 $^{\circ}$C) \citep{Boudjouk1978}. Spectra shown in Fig. \ref{spect_3}.

\section{Laboratory rotational spectroscopy measurements}\label{ftmw}

\begin{figure}
\centering
\includegraphics[angle=0,width=0.5\textwidth]{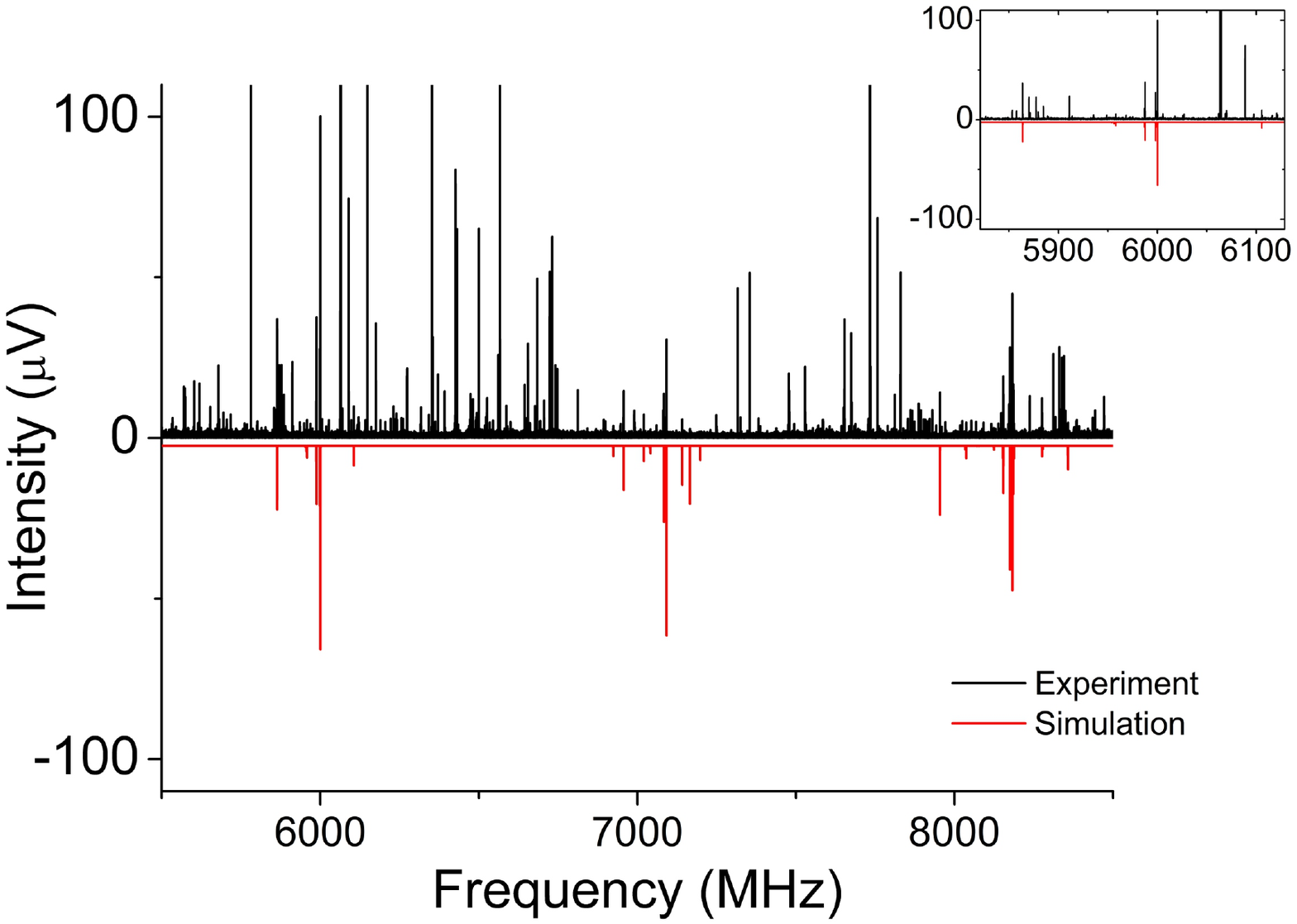}
\caption{Inset of the recorded spectrum arising from gas-phase phenalene. The upper trace is the experimental spectrum, and the red trace is the simulated spectrum derived from the fitted rotational parameters at 1 K for phenalene.} \label{chirped}
\end{figure}

The experimental setup employed to measure the rotational spectra of phenalene consists of broadband chirped-pulse  Fourier-transform microwave (CPFTMW) spectrometer working from 2-18 GHz. The design is largely based on the original design by \citet{Perez2013} with some modifications to allow for direct excitation-detection of the whole 16 GHz bandwidth.
The excitation pulse, ranging from 1 to 9 GHz, is directly generated using an arbitrary waveform generator (Tektronix AWG 70002A, 25 GS/s). This pulse is pre-amplified and frequency-doubled to produce a 4 $\mu$s pulse within the 2-18 GHz range. Subsequently, the pulse is amplified using a solid-state amplifier (SSA Qorvo QPB0220N, 54 dBm output power) that operates over the entire frequency range.
The sample phenalene (see Appendix \ref{synthesis} for description) was placed in a reservoir near the valve orifice and heated to $\sim$80$^{\circ}$C to generate sufficient vapor pressure. These samples were then seeded into a supersonic expansion by mixing with 3.5 bar neon before being expanded into a vacuum chamber. The gas ensemble interacted with the amplified chirped excitation pulse, inducing a macroscopic dipole moment in all polar species within the pulse. The molecular emission was amplified by a 2-18 GHz low-noise amplifier (Miteq LNAS-55-01001800-22-10P) and collected for 20 $\mu$s in the time domain as a free-induction decay using a fast oscilloscope (Tektronix DPO 73304DX, 100 GS/s). For each gas pulse, 20 chirped pulses and a repetition rate of 4 Hz were used, giving an overall repetition rate of 80 Hz. The recorded waveform was then Fourier-transformed to obtain the final frequency-domain spectrum (see Fig. \ref {chirped}).

\section{Initial exploration of ISM compatible formation routes for phenalene} \label{sec:qm}

\begin{figure}
\centering
\includegraphics[angle=0,width=0.5\textwidth]{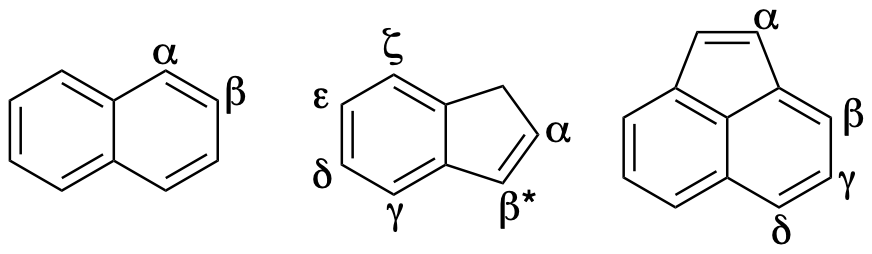}
\caption{Naphthalene (\ce{C10H8}), indene (\ce{C9H8}) and acenaphthylene (\ce{C12H10}) structures along with the symmetry inequivalent positions for reactions with radicals, see text. We note that for indene $\alpha$ and $\beta$ positions yield the same adduct (triangular).} \label{app:qm_neutral}
\end{figure}

To support our conclusions regarding the chemistry of phenalene in the ISM, we performed a preliminary exploration of possible formation routes for it using quantum chemical calculations. As briefly mentioned in the main text, we focused on the formation of phenalene via neutral-neutral reactions involving already detected PAHs: naphthalene (\ce{C10H8}), indene (\ce{C9H8}), and acenaphthylene (\ce{C12H8}), reacting with the arithmetically compatible radicals \ce{C3H3} (linear), \ce{C4H3}, and \ce{CH3} to form \ce{C13H10 + H}. These reactions were investigated using the \textsc{ORCA 6.0} program \citep{neese_software_2022} at the $\omega$B97M-D4/ma-def2-TZVP level of theory \citep{mardirossian_2016,najibi_dft_2020,Caldeweyher2019,Weigend2005,Zheng2011}. All inequivalent positions of \ce{C10H8}, \ce{C9H8}, and \ce{C12H8} were sampled. Although interesting chemical pathways emerged from our investigation, such as the formation of triangular adducts with several of the unsaturated PAH backbones, all the reactions we explored exhibited activation energies that rule them out as viable low-temperature bottom-up formation routes. A summary of the sampled reactions is given in Figure \ref{app:qm_neutral} and Table \ref{app:qm_neutral_table}.

\begin{table}
\caption{Activation energies for neutral-neutral reactions considered in this work, in kJ mol$^{-1}$ with labels from Figure \ref{app:qm_neutral}. All energies are ZPE corrected. }
\label{app:qm_neutral_table}
\centering
\scriptsize
\begin{tabular}{cc}
\hline
\hline
Position & $\Delta U_{a}$ (kJ\,mol$^{-1}$) \\
\hline
\multicolumn{2}{c}{\ce{C10H8} + \ce{C3H3}} \\
\hline
$\alpha$ & 50.8 \\
$\beta$ & 61.6 \\
\hline
\multicolumn{2}{c}{\ce{C9H8} + \ce{C4H3}} \\
\hline
$\alpha^{a}$ & 34.1 \\
$\gamma$ & 61.6\\
$\delta$ & 65.3 \\
$\epsilon$ & 59.0 \\
$\zeta$ & 61.5 \\

\hline
\multicolumn{2}{c}{\ce{C12H8} + \ce{CH3}} \\
\hline
$\alpha$ & 22.8 \\
$\beta$ & 32.3 \\
$\gamma$ & 43.8 \\
$\delta$ & 29.9 \\
\hline
\hline
\end{tabular}
\tablefoot{
 Reference energies are calculated from the separated reactants
\tablefoottext{a} $\alpha$ and $\beta$ positioning of \ce{C4H3} yield the same adduct.}
\end{table}

\begin{figure}
\centering
\includegraphics[angle=0,width=0.50\textwidth]{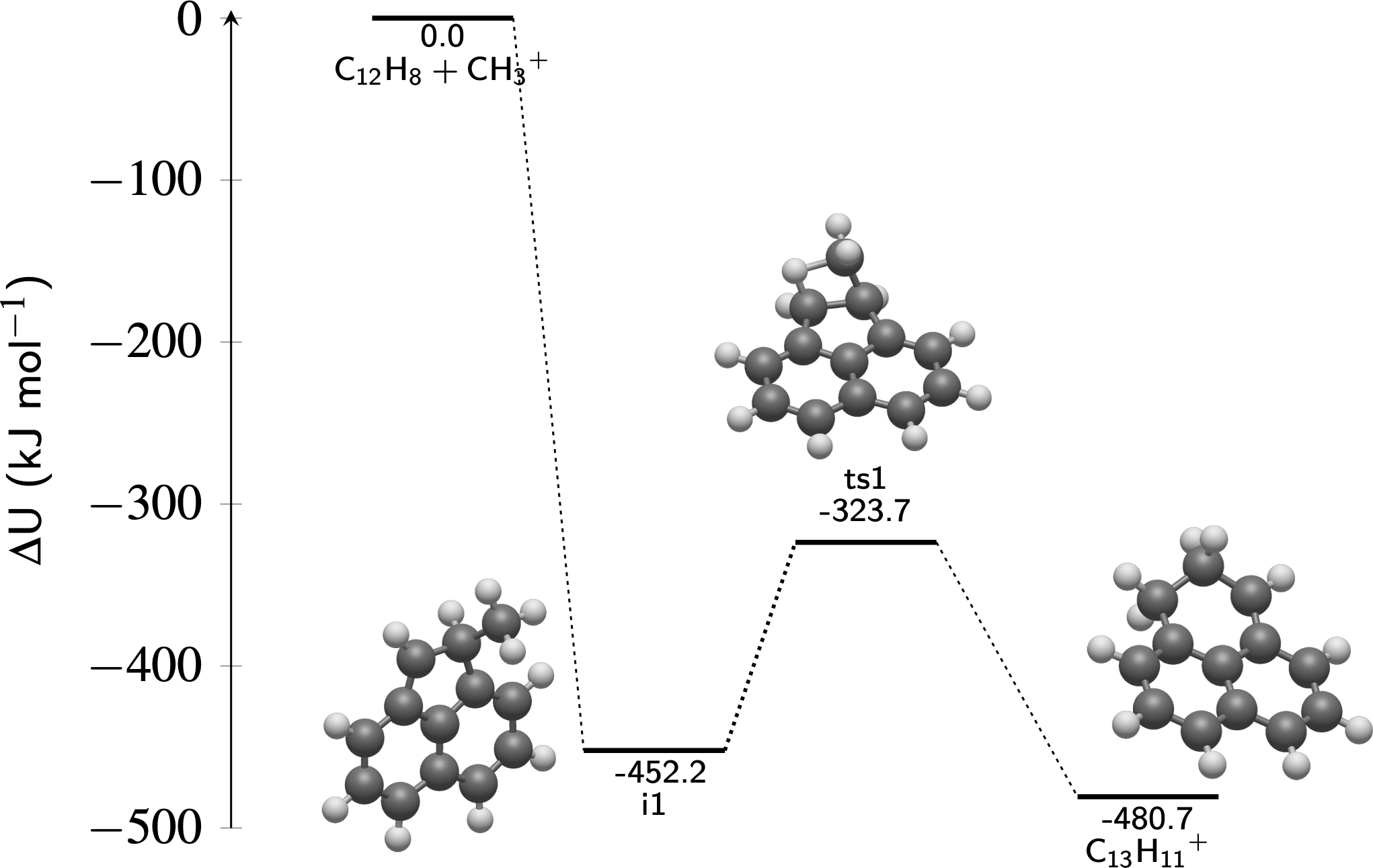}
\caption{Proposed simplified scheme for the formation of \ce{C13H11+}} \label{app:pes}
\end{figure}

In addition to neutral–neutral reactions, we also explored the formation of protonated phenalene (\ce{C13H11+}), which can yield neutral phenalene via dissociative recombination, e.g., \ce{C13H11+ + e- -> C13H10 + H}. However, ion–molecule reactivity is generally less intuitive. To identify viable reaction pathways, we performed a computational retrosynthetic analysis using the single-component artificial force induced reaction (SC-AFIR) method, as implemented in the \textsc{Grrm23} package \citep{maeda_afir_2013,maeda_toward_2023,grrm23}, interfaced with \textsc{Orca}. The calculations employed the GFN2-xTB semiempirical Hamiltonian \citep{Bannwarth2019} and were initiated from the \ce{C13H11+} structure, considering protonation of phenalene adjacent to the \ce{CH2} group. The choice of protonation site was of limited importance, as hydrogen migration emerged as a common pathway in our searches.We conducted six SC-AFIR runs with different model collision energies ({400, 500, 750} kJ mol$^{-1}$), including both global and local searches. The lowest energy search yielded 1245 equilibrium structures and 28 dissociation channels, with comparable results obtained in the replicate runs. Most dissociation pathways involved high-energy, non-viable processes. However, the search also revealed a barrierless and highly exothermic channel involving acenaphthylene (\ce{C12H8}; \citealt{Cernicharo2024}), suggesting a potentially astrochemically relevant route:

\begin{equation}
    \ce{C12H8 + CH3+ -> C13H11+ + h$\nu$} \label{eq:ra}
\end{equation}
The viability of the proposed reaction pathway depends on several conditions that were not addressed in this work and are deferred to a future chemical characterization of these PAHs. While the efficiency of radiative association (Reaction \ref{eq:ra}) is influenced by multiple factors, PAHs represent a favorable case. These relatively large molecules possess deep potential wells and numerous low-energy vibrational modes, which facilitate the dissipation of excess chemical energy through infrared radiation. A recent study on the indenyl cation (\ce{C9H7+}) by \citet{bull_radiative_2025} demonstrates efficient internal relaxation via recurrent fluorescence, suggesting that this mechanism could play an even more prominent role in larger cationic PAHs, such as \ce{C13H11+}. A simplified scheme of the DFT-refined stationary points for the reaction excluding potential bimolecular or competing channels of Reaction \ref{eq:ra} is presented in Figure \ref{app:pes}.

\end{appendix}
\end{document}